# Model-Free Adaptive Predictive Control based Full-Form-Dynamic-Linearization

Feilong Zhang
State Key Laboratory of Robotics, Shenyang Institute of Automation, Chinese Academy of Sciences,
Shenyang 110016, China

*Abstract*—Current model-free adaptive control (MFAC) method lacks robustness to nonlinear systems and its fundamental full-form-dynamic-linearization (FFDL) model only takes one step prediction into consideration. If there exists any time delay in the system, the control law may be invalid. To address this issue, a novel model-free adaptive predictive control (MFAPC) method is developed for a class of discrete-time ingle-input single-output (SISO) nonlinear system. Compared to the current MFAC, this method predicts the outputs of FFDL model over several steps and simultaneously uses more trajectory information in the future. As the result, the proposed method can overcome the time delay problem and is more robust to the choice of parameters in nonlinear system. In addition, we prove that the MFAC is a case of MFAPC when the predictive step $N$=1.

*Index Terms*—full-form-dynamic-linearization, model-free adaptive predictive control, discrete-time single-input single-output nonlinear systems,

## I. INTRODUCTION

A variety of control methods have been proposed and fully realized so far, such as robust control, zero-pole assignment, optimal control, and so on. Most of them are typical model-based control methods in which a priori quantitative or qualitative knowledge of the systems is necessary for the controller design [1]-[5]. However, in most industrial control methods, it is hard to identify the accurate model of a nonlinear system. Hence, the alternative methods such as PID control and generalized predictive control (GPC) have become the most widely used methods in industry settings [6]. Nonetheless, traditional PID control with fixed parameters can hardly meet the control demand of some unknown systems with strong nonlinearities, time-varying parameters and structures. To address this issue, researchers proposed the self-tuning control as a kind of adaptive control, which can adjust the parameters of the PID controller according to the required control performance [7]-[12]. However, the unmodeled dynamics is unfortunately inevitable in the modeling process of adaptive control design approach, which may cause an inherited unsafety in the closed loop control system [13].

In recent years, the data-driven model-free adaptive control (MFAC) has drawn much attention. It has the advantage that the controller design depends on nothing but the measured closed-loop I/O data of the controlled objects. Since the model of system in controller design is not required, the system modeling, the unmodeled dynamics, and the theoretical assumptions on the dynamics of the system do not exist [14]. The off-line model of system is not necessary to build. Instead, MFAC control law is designed through the equivalent dynamic linearization data models at each operating point via the pseudo-gradient (PG) vector. The time-varying PG is based on the deterministic estimation algorithms, merely using the I/O measurement data of the controlled system [14], [15]. In practical applications, MFAC has been successfully implemented in many practical fields, such as: chemical industry [16], [17], linear motor control, injection molding process [18], PH value control [19], and robotic welding process [20]. This is all possible because the MFAC has a simplified discrete control structure, which makes it easier to be implemented through computers.

However, MFAC may lack robustness to some strong nonlinear systems. One reason is that the coefficients of MFAC control law is only determined by the pseudo-gradient (PG) vector in current time. It inherently cannot make full use of the I/O measurement data. In order to improve the stability and robustness of the current MFAC method, we propose the MFAPC, which is a combination of the predictive control and MFAC.

Through the simulation study, we show that the tracking performance, stability and robustness of the MFAPC is better than those of the MFAC. It outperforms the MFAC for two main reasons. Firstly, the MFAPC method can use more future information of the reference trajectory. The system input can be adjusted appropriately before the reference trajectory changes, especially when the operation conditions vary severely in the control systems. Secondly, the index function of the MFAC is only optimal for the error at the current time step, while the index function of the MFAPC takes multiple prediction errors into consideration. To this end, the MFAPC uses more information in the past time to predict the output of the system. Meanwhile, the predictive PG vector $\phi_L(k+i)$ ($i$= 0, 1, 2, ⋯, $N$-1) is estimated by the multi-level hierarchical forecasting method proposed in [26]. This process also uses more past information. These modifications bring several advantages into the MFAPC-controlled system. For some nonlinear unknown systems even with large time delay, the controlled system is faster, more stable and more robust to sudden disturbances which will influence the estimated PG vector. Besides, the key parameter λ of the MFAPC can be chosen from a wider range in simulation, which shares the same advantage of the GPC in industrial settings.

The main contributions of this work are summarized as follows.
1) The FFDL-MFAPC with adjustable step factors is proposed.

With the combination of MFAC and predictive control, the proposed MFAPC has the superior tracking performance than MFAC in simulations. Furthermore, in regard to the relationship between the MFAPC and MFAC, we have an interesting finding: the proposed MFAPC is an elegant extension of the current MFAC, sharing its general structure. Along with this, MFAPC has all the characteristics of the MFAC, whose characteristics are detailed in [14], [21];

2) The BIBO stability and the monotonic convergence of the tracking error dynamics of the FFDL-MFAPC method are firstly analyzed by a contraction mapping based method;

The rest of the paper is organized as follows. In Section II, the equivalent FFDL data predictive model is presented for a class of discrete time nonlinear systems. In Section III, the MFAPC method design and its stability analysis results are presented. In Section IV, the comparison results of simulations are presented to validate the effectiveness and advantages of the proposed MFAPC method. Conclusions are given in Section V. At last, Appendix presents the detailed stability analysis of the proposed method.

## II. DYNAMIC LINEARIZATION DATA PREDICTIVE MODELS FOR DISCRETE-TIME NONLINEAR SYSTEMS

### A. System Model

In this section, the dynamic linearization data modeling method are given as a fundamental tool for the MFAPC controller design, and its basic assumptions, theorem, and insights are given as follows.

The discrete-time SISO nonlinear system is considered as follows:

$$y(k+1) = f(y(k),\cdots,y(k-n_y),u(k),\cdots,u(k-n_u)) \quad (1)$$

where $f(\cdot) \in R$ is an unknown nonlinear function, $n_y$, $n_u \in Z$ are the unknown orders of input $u(k)$ and the output $y(k)$ of the system at time $k$, respectively.

Assume that the nonlinear system (1) conforms with the following assumptions:

*Assumption 1:* The partial derivatives of $f(\cdots)$ with respect to all variables are continuous.

*Assumption 2:* System (1) satisfies generalized Lipschitz condition shown as follows.

$$|y(k_1+1) - y(k_2+1)| \le b\|H(k_1) - H(k_2)\| \quad (2)$$

Where $H(k) = \begin{bmatrix} Y_{L_y}(k) \\ U_{L_u}(k) \end{bmatrix} = [y(k),\cdots,y(k-L_y+1), u(k),\cdots,u(k-L_u+1)]^T$ is a vector that contains control input within a time window $[k-L_u+1,k]$ and output within a moving time window $[k-L_y+1,k]$. Two positive integers $L_y(1 \le L_y \le n_y)$ and $L_u(1 \le L_u \le n_u)$ are called pseudo orders of the system. For more detailed explanations about *Assumption 1* and *Assumption 2* please refer to [14], [21].

*Theorem 1:* For the non-linear system (1) satisfying *Assumptions 1* and *2*, there must exist a time-varying vector $\phi_L(k)$ called PG vector; if $\Delta H(k) \ne 0$, $0 \le L_y$, $1 \le L_u$, system (1) can be transformed into the full-form-dynamic-linearization data model shown as follows

$$\Delta y(k+1) = \phi_L^T(k)\Delta H(k) \quad (3)$$

For any time $k$, we have $\|\phi_L(k)\| \le b$, where

$$\phi_L^T(k) = \begin{bmatrix} \phi_{Ly}(k) \\ \phi_{Lu}(k) \end{bmatrix} = [\phi_1(k),\cdots,\phi_{Ly}(k),\phi_{Ly+1}(k),\cdots,\phi_{Ly+Lu}(k)],$$

$$\Delta H(k) = \begin{bmatrix} \Delta Y_{L_y}(k) \\ \Delta U_{L_u}(k) \end{bmatrix} = [\Delta y(k),\cdots,\Delta y(k-L_y+1), \Delta u(k),\cdots,\Delta u(k-L_u+1)]^T.$$

*Proof:* Refer to [14], [21] for details.

*Remark 1:* Please refer to [14], [21] for the detailed comments and significances about this dynamic linearization data modeling method. [14], [21] also present the relationships between LTI DARMA model and the dynamic linearization data model, and give the suggestions of how to choose the pseudo-orders $L_y$ and $L_u$ of the model.

### B. Predictive System Model

We can rewrite (3) into (4).

$$y(k+1) = y(k) + \phi_L^T(k)\Delta H(k) \quad (4)$$

Here, we define

$$A = \begin{bmatrix} 0 & & & \\ 1 & 0 & & \\ & \ddots & \ddots & \\ & & 1 & 0 \end{bmatrix}_{L_u \times L_u} \quad C = \begin{bmatrix} 0 & & & \\ 1 & 0 & & \\ & \ddots & \ddots & \\ & & 1 & 0 \end{bmatrix}_{L_y \times L_y}$$

$$B^T = \begin{bmatrix} 1 & 0 & \cdots & 0 \end{bmatrix}_{1 \times L_u} \quad D^T = \begin{bmatrix} 1 & 0 & \cdots & 0 \end{bmatrix}_{1 \times L_y}$$

In addition, we define $A^i = 0$ and $C^i = 0$, $i = -1,-2,\cdots$, for the convenience of the following expression. Then, we have finite N step forward prediction equation.

$$\begin{aligned}
\Delta y(k+1) &= \phi_L^T(k)\Delta H(k) = \phi_{Ly}^T(k)\Delta Y(k) + \phi_{Lu}^T(k)\Delta U_L(k) \\
&= \phi_{Ly}^T(k)\Delta Y(k) + \phi_{Lu}^T(k)A\Delta U_L(k-1) + \phi_{Lu}^T(k)B\Delta u(k) \\
\Delta y(k+2) &= \phi_{Ly}^T(k+1)\Delta Y(k+1) + \phi_{Lu}^T(k+1)A\Delta U_L(k) \\
&\quad + \phi_{Lu}^T(k+1)B\Delta u(k+1) \\
\Delta y(k+3) &= \phi_{Ly}^T(k+2)\Delta Y(k+2) + \phi_{Lu}^T(k+2)A\Delta U_L(k+1) \\
&\quad + \phi_{Lu}^T(k+2)B\Delta u(k+2) \\
&\vdots \\
\Delta y(k+N) &= \phi_{Ly}^T(k+N-1)C^{N-1}\Delta Y(k) \\
&\quad + \phi_{Lu}^T(k+N-1)A^N\Delta U_L(k-1) \\
&\quad + \phi_{Ly}^T(k+N-1)C^{N-2}D\Delta y(k+1) + \cdots \\
&\quad + \phi_{Ly}^T(k+N-1)D\Delta y(k+N-1) \\
&\quad + \phi_{Lu}^T(k+N-1)A^{N-1}B\Delta u(k) \\
&\quad + \phi_{Lu}^T(k+N-1)A^{N-2}B\Delta u(k+1) + \cdots \\
&\quad + \phi_{Lu}^T(k+N-1)Bu(k+N-1)
\end{aligned}$$

(5)

Where, $N$ is the predictive step length, $\Delta y(k+i)$ and $\Delta u(k+i)$ are the increment values of the predictive output and the predictive input in the future time $k+i$ ($i=1,2,\cdots,N$), respectively. Here, we define $Y_N(k)$, $\Delta Y_N(k+1)$, $\Delta U_N(k)$, $\Delta U_{Nu}(k)$, $\Psi_Y(k)$, $\Psi_U(k)$, $\tilde{\Psi}_U(k)$, $\Psi_N(k)$ and $\tilde{\Psi}_N(k)$ as follows:

$$Y_N(k+1) = \begin{bmatrix} y(k+1) \\ \vdots \\ y(k+N) \end{bmatrix}_{N\times 1} \quad E = \begin{bmatrix} 1 \\ \vdots \\ 1 \end{bmatrix}_{N\times 1} \quad \Lambda_N = \begin{bmatrix} 1 & & \\ \vdots & \ddots & \\ 1 & \cdots & 1 \end{bmatrix}_{N\times N}$$

$$\Delta U_N(k) = \begin{bmatrix} \Delta u(k) \\ \vdots \\ \Delta u(k+N-1) \end{bmatrix}_{N\times 1} \quad \Delta U_{Nu}(k) = \begin{bmatrix} \Delta u(k) \\ \vdots \\ \Delta u(k+N_u-1) \end{bmatrix}_{Nu\times 1}$$

$$\Delta Y_N(k+1) = Y_N(k+1) - Y_N(k)$$

---

$$\Psi_Y(k) = \begin{bmatrix} \varphi_{11} \\ \varphi_{12} \\ \varphi_{13} \\ \varphi_{14} \\ \vdots \\ \varphi_{1N} \end{bmatrix} = \begin{bmatrix} \phi_{Ly}^T(k) \\ \phi_{Ly}^T(k+1)[C+D\phi_{Ly}^T(k)] \\ \phi_{Ly}^T(k+2)C^2 + \phi_{Ly}^T(k+2)\sum_{i=0}^{1}C^i D\varphi_{12-i} \\ \vdots \\ \phi_{Ly}^T(k+N-1)C^{N-1} \\ + \phi_{Ly}^T(k+N-1)\sum_{i=0}^{N-2}C^i D\varphi_{1N-i-1} \end{bmatrix}_{N\times Ly}$$

$$\Psi_U(k) = \begin{bmatrix} \varphi_{21}^T & \varphi_{22}^T & \varphi_{23}^T & \cdots & \varphi_{2N}^T \end{bmatrix}^T$$

$$= \begin{bmatrix} \phi_{Lu}^T(k)A \\ \phi_{Lu}^T(k+1)A^2 + \phi_{Ly}^T(k+1)D\phi_{Lu}^T(k)A \\ \phi_{Lu}^T(k+2)A^3 + \phi_{Ly}^T(k+2)\sum_{i=0}^{1}C^i D\varphi_{22-i} \\ \vdots \\ \phi_{Lu}^T(k+N-1)A^N + \phi_{Ly}^T(k+N-1)\sum_{i=0}^{N-2}C^i D\varphi_{2N-i-1} \end{bmatrix}$$

$$= [\Psi_{U1}(k), \Psi_{U2}(k), \cdots, \Psi_{ULu-1}(k), 0]_{N\times Lu}$$

$$\Psi_N(k)_{N\times N} = \begin{bmatrix} \psi_{11}, \psi_{12}, \cdots, \psi_{1N} \\ \psi_{21}, \psi_{22}, \cdots, \psi_{2N} \\ \vdots \\ \psi_{N2}, \psi_{N2}, \cdots, \psi_{NN} \end{bmatrix}$$

$$= \begin{bmatrix} \phi_{Lu}^T(k)B & 0 & 0 & \cdots & 0 \\ \phi_{Lu}^T(k+1)AB + \phi_{Ly}^T(k+1)D\phi_{Lu}^T(k)B & \phi_{Lu}^T(k+1)B & 0 & \cdots & 0 \\ \phi_{Lu}^T(k+2)A^2 B + \phi_{Ly}^T(k+2)\sum_{i=0}^{1}C^i D\psi_{2-i,1} & \phi_{Lu}^T(k+2)AB + \phi_{Ly}^T(k+2)D\phi_{Lu}^T(k+1)B & \phi_{Lu}^T(k+2)B & \cdots & 0 \\ \vdots & \vdots & \vdots & \ddots & \vdots \\ \phi_{Lu}^T(k+N-1)A^{N-1}B + \phi_{Ly}^T(k+N-1)\sum_{i=0}^{N-2}C^i D\psi_{N-i-1,1} & \phi_{Lu}^T(k+N-1)A^{N-2}B + \phi_{Ly}^T(k+N-1)\sum_{i=0}^{N-3}C^i D\psi_{N-i-1,2} & \phi_{Lu}^T(k+N-1)A^{N-3}B + \phi_{Ly}^T(k+N-1)\sum_{i=0}^{N-4}C^i D\psi_{N-i-1,3} & \cdots & \phi_{Lu}^T(k+N-1)B \end{bmatrix}$$

$$\tilde{\Psi}_N(k)_{N\times N} = \Lambda_N \Psi_N(k)$$

$$= \begin{bmatrix} \phi_{Lu}^T(k)B & 0 & 0 & \cdots & 0 \\ \phi_{Lu}^T(k+1)AB + \phi_{Lu}^T(k)B + \phi_{Ly}^T(k+1)D\phi_{Lu}^T(k)B & \phi_{Lu}^T(k+1)B & 0 & \cdots & 0 \\ \sum_{j=1}^{3}[\phi_{Lu}^T(k+j-1)A^{j-1}B + \phi_{Ly}^T(k+j-1)\sum_{i=0}^{j-2}C^i D\psi_{j-i-1,1}] & \phi_{Lu}^T(k+2)AB + \phi_{Lu}^T(k+1)B + \phi_{Ly}^T(k+2)D\phi_{Lu}^T(k+1)B & \phi_{Lu}^T(k+2)B & \cdots & 0 \\ \vdots & \vdots & \vdots & \ddots & \vdots \\ \sum_{j=1}^{N}[\phi_{Lu}^T(k+j-1)A^{j-1}B + \phi_{Ly}^T(k+j-1)\sum_{i=0}^{j-2}C^i D\psi_{j-i-1,1}] & \sum_{j=2}^{N}[\phi_{Lu}^T(k+j-1)A^{j-2}B + \phi_{Ly}^T(k+j-1)\sum_{i=0}^{j-3}C^i D\psi_{j-i-1,2}] & \sum_{j=3}^{N}[\phi_{Lu}^T(k+j-1)A^{j-3}B + \phi_{Ly}^T(k+j-1)\sum_{i=0}^{j-4}C^i D\psi_{j-i-1,3}] & \cdots & \phi_{Lu}^T(k+N-1)B \end{bmatrix}$$

$$\tilde{\boldsymbol{\Psi}}_Y(k) = \boldsymbol{\Lambda}_N \boldsymbol{\Psi}_Y(k) = \begin{bmatrix} \boldsymbol{\phi}_{Ly}^T(k) \\ \boldsymbol{\phi}_{Ly}^T(k+1)[\boldsymbol{C} + \boldsymbol{D}\boldsymbol{\phi}_{Ly}^T(k)] + \boldsymbol{\phi}_{Ly}^T(k) \\ \vdots \\ \sum_{j=1}^{N} [\boldsymbol{\phi}_{Ly}^T(k+j-1)\boldsymbol{C}^{j-1} \\ + \boldsymbol{\phi}_{Ly}^T(k+j-1)\sum_{i=0}^{j-2} \boldsymbol{C}^i \boldsymbol{D}\boldsymbol{\varphi}_{1 j-i-1}] \end{bmatrix}$$

$$= [\tilde{\boldsymbol{\Psi}}_{Y1}(k), \tilde{\boldsymbol{\Psi}}_{Y2}(k), \cdots, \tilde{\boldsymbol{\Psi}}_{YLy}(k)]_{N \times Ly}$$

$$\tilde{\boldsymbol{\Psi}}_U(k) = \boldsymbol{\Lambda}_N \boldsymbol{\Psi}_U(k)$$

$$= \begin{bmatrix} \boldsymbol{\phi}_{Lu}^T(k)\boldsymbol{A} \\ \boldsymbol{\phi}_{Lu}^T(k+1)\boldsymbol{A}^2 + \boldsymbol{\phi}_{Ly}^T(k+1)\boldsymbol{D}\boldsymbol{\phi}_{Lu}^T(k)\boldsymbol{A} + \boldsymbol{\phi}_{Lu}^T(k)\boldsymbol{A} \\ \sum_{j=1}^{3}[\boldsymbol{\phi}_{Lu}^T(k+j-1)\boldsymbol{A}^j + \boldsymbol{\phi}_{Ly}^T(k+j-1)\sum_{i=0}^{j-2}\boldsymbol{C}^i\boldsymbol{D}\boldsymbol{\varphi}_{2 j-i-1}] \\ \vdots \\ \sum_{j=1}^{N}[\boldsymbol{\phi}_{Lu}^T(k+j-1)\boldsymbol{A}^j + \boldsymbol{\phi}_{Ly}^T(k+j-1)\sum_{i=0}^{j-2}\boldsymbol{C}^i\boldsymbol{D}\boldsymbol{\varphi}_{2 j-i-1}] \end{bmatrix}$$

$$= \left[\tilde{\boldsymbol{\Psi}}_{U1}(k), \tilde{\boldsymbol{\Psi}}_{U2}(k), \cdots, \tilde{\boldsymbol{\Psi}}_{ULu-1}(k), 0\right]_{N \times Lu}$$

---

where, $\varphi_{1i}$ is the $i$-th row of the $\boldsymbol{\Psi}_Y(k)$; $\varphi_{2i}$ is the $i$-th row of the $\boldsymbol{\Psi}_U(k)$, and $\boldsymbol{\Psi}_{Uj}(k)$ is the $j$-th column of the $\boldsymbol{\Psi}_U(k)$; $\psi_{ij}$ is the element in the $i$-th row and $j$-th column of $\boldsymbol{\Psi}_N(k)$; $\tilde{\boldsymbol{\Psi}}_{Yj}(k)$ is the $j$-th column of the $\tilde{\boldsymbol{\Psi}}_Y(k)$; $\tilde{\boldsymbol{\Psi}}_{Uj}(k)$ is the $j$-th column of the $\tilde{\boldsymbol{\Psi}}_{Uj}(k)$; In addition, we define $\varphi_{1i} = 0$, $\varphi_{2i} = 0$ and $\psi_{ij} = 0$, $i = -1, -2, \cdots$.

Then, the prediction equation (5) can be written as (6):

$$\Delta \boldsymbol{Y}_N(k+1) = \boldsymbol{\Psi}_Y(k)\Delta \boldsymbol{Y}_{Ly}(k) + \boldsymbol{\Psi}_U(k)\Delta \boldsymbol{U}_{Lu}(k-1) + \boldsymbol{\Psi}_N(k)\Delta \boldsymbol{U}_N(k) \quad (6)$$

Furthermore, both sides of equation (6) are left multiplied by $\boldsymbol{\Lambda}_N$, then (6) can be rewritten as:

$$\boldsymbol{Y}_N(k+1) = \boldsymbol{E}y(k) + \boldsymbol{\Lambda}_N \boldsymbol{\Psi}_Y(k)\Delta \boldsymbol{Y}_{Ly}(k) + \boldsymbol{\Lambda}_N \boldsymbol{\Psi}_U(k)\Delta \boldsymbol{U}_{Lu}(k-1)$$
$$+ \boldsymbol{\Lambda}_N \boldsymbol{\Psi}_N(k)\Delta \boldsymbol{U}_N(k)$$
$$= \boldsymbol{E}y(k) + \tilde{\boldsymbol{\Psi}}_Y(k)\Delta \boldsymbol{Y}_{Ly}(k) + \tilde{\boldsymbol{\Psi}}_U(k)\Delta \boldsymbol{U}_{Lu}(k-1)$$
$$+ \tilde{\boldsymbol{\Psi}}_N(k)\Delta \boldsymbol{U}_N(k) \quad (7)$$

Define $N_u$ as control step length. If $\Delta u(k+j-1) = 0$, $N_u < j \le N$, the equation (7) can be rewritten into

$$\boldsymbol{Y}_N(k+1) = \boldsymbol{E}y(k) + \tilde{\boldsymbol{\Psi}}_Y(k)\Delta \boldsymbol{Y}_{Ly}(k) + \tilde{\boldsymbol{\Psi}}_U(k)\Delta \boldsymbol{U}_{Lu}(k-1)$$
$$+ \tilde{\boldsymbol{\Psi}}_{Nu}(k)\Delta \boldsymbol{U}_{Nu}(k) \quad (8)$$

where $\tilde{\boldsymbol{\Psi}}_{Nu}(k)$ is defined as follow.

---

$$\tilde{\boldsymbol{\Psi}}_{Nu}(k)_{N \times Nu}$$

$$= \begin{bmatrix} \boldsymbol{\phi}_{Lu}^T(k)\boldsymbol{B} & 0 & \cdots & 0 \\ \boldsymbol{\phi}_{Lu}^T(k+1)\boldsymbol{AB} + \boldsymbol{\phi}_{Ly}^T(k+1)\boldsymbol{D}\boldsymbol{\phi}_{Lu}^T(k)\boldsymbol{B} + \boldsymbol{\phi}_{Lu}^T(k)\boldsymbol{B} & \boldsymbol{\phi}_{Lu}^T(k+1)\boldsymbol{B} & \cdots & 0 \\ \vdots & \vdots & \ddots & \vdots \\ \sum_{j=1}^{Nu}[\boldsymbol{\phi}_{Lu}^T(k+j-1)\boldsymbol{A}^{j-1}\boldsymbol{B} + \boldsymbol{\phi}_{Ly}^T(k+j-1)\sum_{i=0}^{j-2}\boldsymbol{C}^i\boldsymbol{D}\psi_{j-i-1,1}] & \sum_{j=2}^{Nu}[\boldsymbol{\phi}_{Lu}^T(k+j-1)\boldsymbol{A}^{j-2}\boldsymbol{B} + \boldsymbol{\phi}_{Ly}^T(k+j-1)\sum_{i=0}^{j-3}\boldsymbol{C}^i\boldsymbol{D}\psi_{j-i-1,2}] & \cdots & \boldsymbol{\phi}_{Lu}^T(k+N_u-1)\boldsymbol{B} \\ \vdots & \vdots & \vdots & \vdots \\ \sum_{j=1}^{N}[\boldsymbol{\phi}_{Lu}^T(k+j-1)\boldsymbol{A}^{j-1}\boldsymbol{B} + \boldsymbol{\phi}_{Ly}^T(k+j-1)\sum_{i=0}^{j-2}\boldsymbol{C}^i\boldsymbol{D}\psi_{j-i-1,1}] & \sum_{j=2}^{N}[\boldsymbol{\phi}_{Lu}^T(k+j-1)\boldsymbol{A}^{j-2}\boldsymbol{B} + \boldsymbol{\phi}_{Ly}^T(k+j-1)\sum_{i=0}^{j-3}\boldsymbol{C}^i\boldsymbol{D}\psi_{j-i-1,2}] & \cdots & \sum_{j=Nu}^{N}[\boldsymbol{\phi}_{Lu}^T(k+j-1)\boldsymbol{A}^{j-Nu}\boldsymbol{B} + \boldsymbol{\phi}_{Ly}^T(k+j-1)\sum_{i=0}^{j-Nu-1}\boldsymbol{C}^i\boldsymbol{D}\psi_{j-i-1,Nu}] \end{bmatrix}$$

## III. MODEL-FREE ADAPTIVE PREDICTIVE CONTROL DESIGN AND STABILITY ANALYSIS

In this section, the design of MFAPC method will firstly be presented. Based on the finding in [14] that introduced the relationships between the MFAC, the traditional adaptive control, and the well-known PID as well as the controller parameters choosing suggestions, we present some possible relationships among the MFAPC, MFAC, MFAPC-PID and the MFAC-PID. Then, we present the stability analysis with some necessary Theorems and Lemma.

### A. Design of Model Free Adaptive Predictive Control

A weighted control input cost function is shown below:

$$J = \left[\boldsymbol{Y}_N^*(k+1) - \boldsymbol{Y}_N(k+1)\right]^T \left[\boldsymbol{Y}_N^*(k+1) - \boldsymbol{Y}_N(k+1)\right] + \lambda \Delta \boldsymbol{U}_{Nu}^T(k) \Delta \boldsymbol{U}_{Nu}(k) \quad (9)$$

Where, $\lambda$ is the weighted constant; $\tilde{\boldsymbol{Y}}_N^*(k+1) = \left[y^*(k+1), \cdots, y^*(k+N)\right]^T$ is the desired system output signal vector, where $y^*(k+i)$ is the prediction of the future output at the time $(k+i)$ ($i = 1, 2, \cdots, N$).

Substitute Equation (8) into Equation (9) and solve the optimization condition $\partial J/\partial \Delta \tilde{U}_{N_u}(k) = 0$, we have:

$$\Delta U_{Nu}(k) = [\tilde{\boldsymbol{\Psi}}_{Nu}^T(k)\tilde{\boldsymbol{\Psi}}_{Nu}(k) + \lambda \boldsymbol{I}]^{-1}\tilde{\boldsymbol{\Psi}}_{Nu}^T(k)[\rho_{Ly+1}(\boldsymbol{Y}_N^*(k+1)$$
$$-\boldsymbol{E}y(k)) - \tilde{\boldsymbol{\Psi}}_Y(k)\Lambda_Y \Delta \boldsymbol{Y}_{Ly}(k) - \tilde{\boldsymbol{\Psi}}_U(k)\Lambda_U \Delta \boldsymbol{U}_{Lu}(k-1)] \quad (10)$$

where the adjustable step factors $\Lambda_Y = diag[\rho_1, \cdots, \rho_{Ly}]$, $\rho_{Ly+1}$ and $\Lambda_U = diag[\rho_{Ly+2}, \cdots, \rho_{Ly+Lu+1}]$ are introduced to make the controller algorithm more flexible and analysis the stability of the system, $\rho_i < 1$ $(i=1,2,\cdots,L_y+L_u)$. The current input is given by

$$u(k) = u(k-1) + \boldsymbol{g}^T \Delta \boldsymbol{U}_{Nu}(k) \quad (11)$$

where $\boldsymbol{g} = [1,0,\cdots,0]^T$.

*Remark 2:* $\tilde{\boldsymbol{\Psi}}_{Nu}(k)$, $\tilde{\boldsymbol{\Psi}}_Y(k)$ and $\tilde{\boldsymbol{\Psi}}_U(k)$ in the Equation (10) contain the unknown $\boldsymbol{\phi}_{Ly}^T(k+i)$ and $\boldsymbol{\phi}_{Lu}^T(k+i)$ ($i= 0, 1, 2, \cdots, N$-1) which need to be replaced by their estimated and predicted values $\hat{\boldsymbol{\phi}}_{Ly}^T(k+i)$ and $\hat{\boldsymbol{\phi}}_{Lu}^T(k+i)$. The $\hat{\boldsymbol{\phi}}_{Ly}^T(k)$ and $\hat{\boldsymbol{\phi}}_{Lu}^T(k)$ are estimated by the projection algorithm in [14][21]. The $\boldsymbol{\phi}_{Ly}^T(k+i)$ and $\boldsymbol{\phi}_{Lu}^T(k+i)$, $i= 1, 2, \cdots, N$-1 are predicted by the data-driven multi-level hierarchical forecasting method proposed in [21], [24]-[27]. From these references, we know that the $\hat{\boldsymbol{\phi}}_L(k+i)$ ($i= 0, 1, 2, \cdots, N$-1), which are the linear combination of the $\hat{\boldsymbol{\phi}}_L(k)$, $\hat{\boldsymbol{\phi}}_L(k-1)$, $\cdots$, $\hat{\boldsymbol{\phi}}_L(k-n_p+1)$, are bounded. Let us define $\hat{\tilde{\boldsymbol{\Psi}}}_{Nu}(k)$, $\hat{\tilde{\boldsymbol{\Psi}}}_Y(k)$ and $\hat{\tilde{\boldsymbol{\Psi}}}_U(k)$ as the estimated matrixes of the $\tilde{\boldsymbol{\Psi}}_{Nu}(k)$, $\tilde{\boldsymbol{\Psi}}_Y(k)$ and $\tilde{\boldsymbol{\Psi}}_U(k)$, respectively. Then, according to the definition of the norms of matrix, the norms of $\hat{\tilde{\boldsymbol{\Psi}}}_{Nu}(k)$, $\hat{\tilde{\boldsymbol{\Psi}}}_Y(k)$ and $\hat{\tilde{\boldsymbol{\Psi}}}_U(k)$ are bounded.

Then we get the proposed MFAPC control input (12)

$$\Delta U_{Nu}(k) = [\hat{\tilde{\boldsymbol{\Psi}}}_{Nu}^T(k)\hat{\tilde{\boldsymbol{\Psi}}}_{Nu}(k) + \lambda \boldsymbol{I}]^{-1}\hat{\tilde{\boldsymbol{\Psi}}}_{Nu}^T(k)[\rho_{Ly+1}(\boldsymbol{Y}_N^*(k+1)$$
$$-\boldsymbol{E}y(k)) - \hat{\tilde{\boldsymbol{\Psi}}}_Y(k)\Lambda_Y \Delta \boldsymbol{Y}_{Ly}(k) - \hat{\tilde{\boldsymbol{\Psi}}}_U(k)\Lambda_U \Delta \boldsymbol{U}_{Lu}(k-1)] \quad (12)$$

The current control law is

$$u(k) = u(k-1) + \boldsymbol{g}^T \Delta \boldsymbol{U}_{Nu}(k) \quad (13)$$

*Remark 3:* The method of how to choose pseudo orders $L_y$, $L_u$ of the data model are detailed in [14], [21]. In practical experiments, we'd better first try the relatively small values of the pseudo orders $L_y$, $L_u$, then tune the other parameters $\lambda$, $\mu$, $\eta$, $\rho_i$, $i=1,\cdots, L_y+L_u$, among which the $\lambda$ plays a major role in stability analysis and should be tuned firstly to guarantee the stability of the system. If it does not converge well, the higher pseudo orders $L_y$, $L_u$ need be adopted. Then we repeat the above process.

We typically choose the sufficiently large predictive step length $N$ which should be larger than the time-delay or make the dynamics of the system to be covered. The larger $N$ may improve the robustness of the system. However, this may degenerate the transient and tracking performance and increase the online computational burden. For $N_u$, while the larger control horizon $N_u$ may increase the sensitivity and tracking ability of the system, it may cause the degradation of stability and robustness of the system, and online computational cost will be increased inevitably for matrix dimension extended. $N_u$ can be chosen to be 1 for simple systems (e.g. low-order linear system), whereas for complex systems, a larger value of $N_u$ can improve the transient and tracking performance. When $N_u=1$, $\hat{\tilde{\boldsymbol{\Psi}}}_{Nu}(k)$ will become a column vector, and it will reduce the computational cost and computational time.

*Remark 4:* The following cases are given as the special cases of the proposed MFAPC method.

Case 1: When $N_u = 1$, we have the following simplified control output (14), which does not have the inverse calculation of matrix

$$\Delta U_{Nu}(k) = \frac{1}{[\hat{\tilde{\boldsymbol{\Psi}}}_{Nu}^T(k)]_{1\times N}[\hat{\tilde{\boldsymbol{\Psi}}}_{Nu}(k)]_{N\times 1} + \lambda}$$
$$\bullet [\hat{\tilde{\boldsymbol{\Psi}}}_{Nu}^T(k)]_{1\times N}[\rho_{Ly+1}(\boldsymbol{Y}_N^*(k+1) - \boldsymbol{E}y(k)) - [\hat{\tilde{\boldsymbol{\Psi}}}_Y(k)]_{N\times L_y}$$
$$\begin{bmatrix} \rho_1 & & \\ & \ddots & \\ & & \rho_{Ly} \end{bmatrix} \begin{bmatrix} \Delta y(k) \\ \vdots \\ \Delta y(k-L_y+1) \end{bmatrix}$$
$$- [\hat{\tilde{\boldsymbol{\Psi}}}_U(k)]_{N\times L_u} \begin{bmatrix} \rho_{Ly+2} & & \\ & \ddots & \\ & & \rho_{Ly+Lu+1} \end{bmatrix} \begin{bmatrix} \Delta u(k-1) \\ \vdots \\ \Delta u(k-L_u) \end{bmatrix}] \quad (14)$$

Case 2: When $L_y = 2$ and $L_u = 1$, we have the control output:

$$\Delta u(k) = K_P \Delta e(k) + K_I \begin{bmatrix} y^*(k+1) - y(k) \\ \vdots \\ y^*(k+N) - y(k) \end{bmatrix} \quad (15)$$
$$+ K_D[\Delta e(k) - \Delta e(k-1)]$$

where

$$K_P = \rho_1 \boldsymbol{g}^T[\hat{\tilde{\boldsymbol{\Psi}}}_{Nu}^T(k)\hat{\tilde{\boldsymbol{\Psi}}}_{Nu}(k) + \lambda \boldsymbol{I}]^{-1}\hat{\tilde{\boldsymbol{\Psi}}}_{Nu}^T(k)[\hat{\tilde{\boldsymbol{\Psi}}}_Y(k)]_{N\times 2:1}$$
$$+ \rho_2 \boldsymbol{g}^T[\hat{\tilde{\boldsymbol{\Psi}}}_{Nu}^T(k)\hat{\tilde{\boldsymbol{\Psi}}}_{Nu}(k) + \lambda \boldsymbol{I}]^{-1}\hat{\tilde{\boldsymbol{\Psi}}}_{Nu}^T(k)[\hat{\tilde{\boldsymbol{\Psi}}}_Y(k)]_{N\times 2:2},$$

$$K_I = \rho_3 \boldsymbol{g}^T[\hat{\tilde{\boldsymbol{\Psi}}}_{Nu}^T(k)\hat{\tilde{\boldsymbol{\Psi}}}_{Nu}(k) + \lambda \boldsymbol{I}]^{-1}\hat{\tilde{\boldsymbol{\Psi}}}_{Nu}^T(k),$$

$$K_D = -\rho_2 \boldsymbol{g}^T[\hat{\tilde{\boldsymbol{\Psi}}}_{Nu}^T(k)\hat{\tilde{\boldsymbol{\Psi}}}_{Nu}(k) + \lambda \boldsymbol{I}]^{-1}\hat{\tilde{\boldsymbol{\Psi}}}_{Nu}^T(k)[\hat{\tilde{\boldsymbol{\Psi}}}_Y(k)]_{N\times 2:2},$$

$e(k) = y_d - y(k)$, and we rewrite

$[\hat{\tilde{\boldsymbol{\Psi}}}_Y(k)]_{N\times 2} := -[[\hat{\tilde{\boldsymbol{\Psi}}}_Y(k)]_{N\times 2:1}, [\hat{\tilde{\boldsymbol{\Psi}}}_Y(k)]_{N\times 2:2}]$,

(15) obviously represents the PID form of MFAPC.

Furthermore, when $N = 1$ and the corresponding $N_u = 1$, we have

$$\Delta u(k) = \frac{\hat{\phi}_3(k)}{\lambda + |\hat{\phi}_3(k)|^2}[\rho_3(y^* - y(k))$$
$$- [\hat{\phi}_1(k) \quad \hat{\phi}_2(k)]_{1\times 2} \begin{bmatrix} \rho_1 & \\ & \rho_2 \end{bmatrix} \begin{bmatrix} \Delta y(k) \\ \Delta y(k-1) \end{bmatrix}] \quad (16)$$
$$= K_P \Delta e(k) + K_I e(k) + K_D[\Delta e(k) - \Delta e(k-1)]$$

where

$$K_P = \frac{\rho_1 \hat{\phi}_3(k)\hat{\phi}_1(k)}{\lambda + |\hat{\phi}_3(k)|^2} + \frac{\rho_2 \hat{\phi}_3(k)\hat{\phi}_2(k)}{\lambda + |\hat{\phi}_3(k)|^2}, \quad K_I = \frac{\rho_3 \hat{\phi}_3(k)}{\lambda + |\hat{\phi}_3(k)|^2},$$

$$K_D = \frac{-\rho_2 \hat{\phi}_3(k)\hat{\phi}_2(k)}{\lambda + |\hat{\phi}_3(k)|^2}$$

[14] claims that (16) is the well-known PID controller structure, which belongs to MFAC structure thereof.

Case 3: When $N=1$ and the corresponding $N_u=1$, the MFAPC degenerates into the MFAC shown as (17)

$$\Delta u(k) = \frac{\rho_{Ly+1}}{\lambda + |\hat{\phi}_{Ly+1}(k)|^2} \hat{\phi}_{Ly+1}(k)[\rho_{Ly+1}(y^* - y(k))$$

$$-[\hat{\phi}_1(k) \cdots \hat{\phi}_{Ly}(k)] \begin{bmatrix} \rho_1 & & \\ & \ddots & \\ & & \rho_{Ly} \end{bmatrix} \begin{bmatrix} \Delta y(k) \\ \vdots \\ \Delta y(k - L_y + 1) \end{bmatrix}$$

$$- \begin{bmatrix} \hat{\phi}_{Lu+2}(k) \\ \vdots \\ \hat{\phi}_{Lu+Lu}(k) \end{bmatrix}^T \begin{bmatrix} \rho_{Ly+2} & & \\ & \ddots & \\ & & \rho_{Ly+Lu} \end{bmatrix} \begin{bmatrix} \Delta u(k-1) \\ \vdots \\ \Delta u(k-L_u+1) \end{bmatrix}]$$

(17)

From Case 3, we can conclude that the proposed MFAPC degrades to the MFAC when the predictive step N=1. In addition, Fig. 1 shows the relationships among MFAPC, MFAC, the PID in MFAC (MFAC-PID), and the predictive PID in MFAPC (MFAPC-PID).

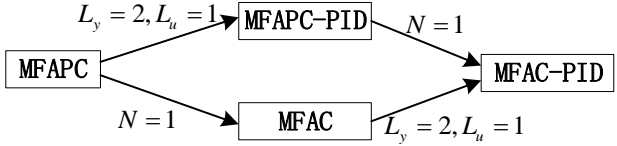

Fig. 1 The relationships among MFAPC, MFAC, MFAC-PID and MFAPC- PID

B. *Stability Analysis of MFAPC*

This section gives some Lemmas, assumptions, and the proof of stability of MFAPC.

*Lemma 1* ([28]): Let $A = \begin{bmatrix} a_1 & \cdots & \cdots & a_{L-1} & a_L \\ 1 & 0 & \cdots & 0 & 0 \\ 0 & 1 & \cdots & 0 & 0 \\ \vdots & \vdots & \vdots & \vdots & \vdots \\ 0 & 0 & \cdots & 1 & 0 \end{bmatrix}$. If

$\sum_{i=1}^{L} |a_i| < 1$, then $\sigma(A) < 1$, where $\sigma(A)$ is the spectral radius of $A$.

*Lemma 2*: ([29]) Given $A \in R^{n \times n}$, for any given $\varepsilon > 0$, there exists an induced consistent matrix norm such that

$$\|A\|_v \leq \sigma(A) + \varepsilon$$

where $\rho(A)$ is the spectral radius of $A$.

*Assumption 3*: We quota *Assumption 3* and *Assumption 4* in [14] to save the room for this paper.

*Theorem 2*: If the system is described by (1) and controlled by the MFAPC method (12)-(13) with the desired trajectory $y_d(k) = y_d = const$, there exists a $\lambda_{\min}$, such that, when $\lambda > \lambda_{\min}$, it guarantees: 1) $\lim_{k \to \infty} |y(k+1) - y^*| = 0$; 2) the control system is BIBO stability.

*Proof*: For completeness and compactness of the proposed method, the Appendix presents the proof of *Theorem 2*, which is inspired by [14], [21], [30].

## IV. SIMULATIONS

Example 2: A number of examples are given in [14], [32] to show the effectiveness and the advantages of MFAC methods by comparing with other typical DDC methods, data-driven PID (DD-PID), iterative feedback tuning (IFT), and virtual reference feedback tuning (VRFT), respectively. The conclusion in [14] is that the tracking performance of Hou's MFAC-PI is better than the above DDC method in its simulation. In this example, comparisons of the simulation results between MFAPC-PI and MFAC-PI are given under the same model which is unknown to the controller design process and is from [14], [32]:

$$y(k+1) = 0.6 y(k) - 0.1 y(k-1) + 1.8 u(k) - 1.8 u^2(k) + 0.6 u^3(k)$$
$$- 0.15 u(k-1) + 0.15 u^2(k-1) - 0.05 u^3(k-1)$$

(18)

The model is merely applied to generate output data for MFAPC and MFAC. The desired trajectory is the same as example 1, and it is more difficult to be tracked compared with Hou's [14], [32]. The initial values of MFAPC and MFAC are identical. The parameter settings for MFAPC-PI and MFAC-PI methods are given in Table III. The control output of MFAPC-PI is written in the following form:

$$\Delta u(k) = \mathbf{g}^T [\hat{\mathbf{\Psi}}_{Nu}^T(k) \hat{\mathbf{\Psi}}_{Nu}(k) + \lambda \mathbf{I}]^{-1} \hat{\mathbf{\Psi}}_{Nu}^T(k) \bullet$$
$$[\rho_2 (\mathbf{Y}_N^*(k+1) - \mathbf{E} y(k)) - \rho_1 \hat{\mathbf{\Psi}}_Y(k) \Delta y(k)]$$

(19)

TABLE III Parameter Settings for MFAC-PI and MFAPC-PI

| Parameter | MFAC-PI in [14] | MFAC-PI | MFAPC-PI | MFAPC2-PI |
|---|---|---|---|---|
| Order | $Ly=1, Lu=1$ | $Ly=1, Lu=1$ | $Ly=1, Lu=1$ | $Ly=1, Lu=1$ |
| $\lambda$ | 1 | 0.6 (0-0.5 is not stable), 1, 2, 4, 10, 20 | 2 | 2 |
| $\rho$ | [0.4, 0.4] | [0.4, 0.4] | [0.4, 0.4] | [0.4, 0.4] |
| $\mu, \eta$ | 1, 1 | 10, 0.5 | 10, 0.5 | 10, 0.5 |
| Initial value | [0.1, 0.1] (By all means) | [0.1, 0.1] | [0.1, 0.1] | [0.1, 0.1] |
| Reset value | Null (By all means) | Null | Null | Null |
| $(u(0), u(1), u(2))$ | (0, 0, 0) | (0, 0, 0) | (0, 0, 0) | (0, 0, 0) |
| $(y(0), y(1), y(2))$ | (0, 0, 0) | (0, 0, 0) | (0, 0, 0) | (0, 0, 0) |
| Predictive step | No choice | No choice | N=3 | N=2 |
| Control step | No choice) | No choice | Nu=3 | Nu=2 |

In [14], it was concluded that the penalty factor $\lambda$ plays a major role in the stability analysis and in applications. In Fig. 2, the MFAC-PI is applied with different values of $\lambda$ in order to make comparisons with MFAPC-PI applied with $\lambda=2$. In addition, Fig.

2 shows the tricking performance of the system when the MFAPC-PI is applied with different values of λ.

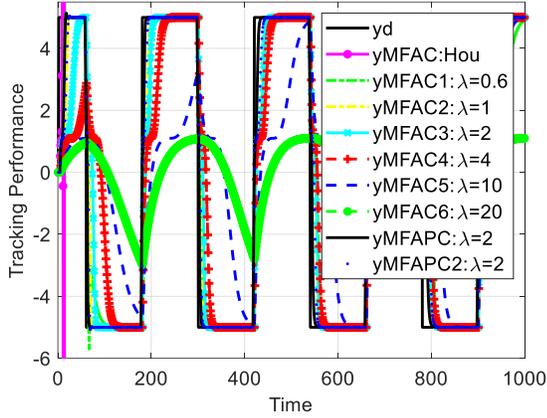

Fig. 1 Tracking performance

The comments of the simulation are given as follows:
1) In Fig. 1, the output of the system controlled by MFAC-PI with the parameters in [14] cannot converge to the desired trajectory until estimate algorithm parameters μ and η are changed as in MFAC-PI in TABLE II.
2) In Fig. 1, the smaller λ is applied in MFAC, resulting in the better tracking performances. When the λ is in [0, 0.5], the system will be unstable. Therefore, we choose the λ=0.6 as the smallest parameter of the MFAC in this simulation.
3) In Fig. 1, the tracking performance of MFAPC with predictive step N=3 is better than that of MFAPC2 with N=2, and both are much better than the tracking performances of MFAC in this example.
4) According to the comparison between Fig. 1 and Fig. 2, we can see that when λ increases, the tracking performances of the system controlled by MFAC obviously becomes worse. When λ=20, the output cannot track the desired trajectory, as shown in Fig. 1. As shown in Fig. 2, the system controlled by MFAPC can track the desired trajectory, even when λ=20. Moreover, its tracking performance is better than that of MFAC with λ=10. Thus, it validates the claim that MFAPC is not sensitive and has stronger robustness to the change of the key parameter λ compared with MFAC in this example, and we can have a wider range for choosing the λ in practice.

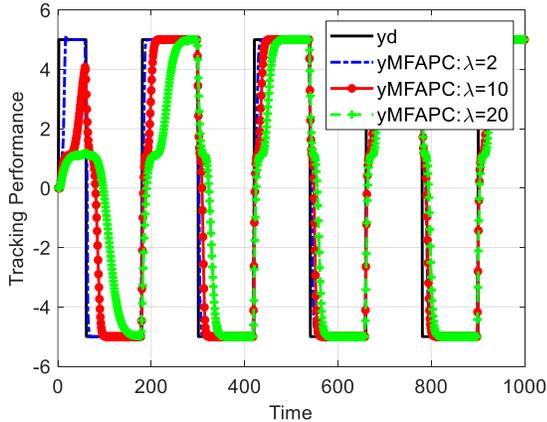

Fig. 2 Tracking performance

Fig. 3 shows the components of the PG estimation of MFAPC applied with λ=2 and that of MFAC applied with λ=2.

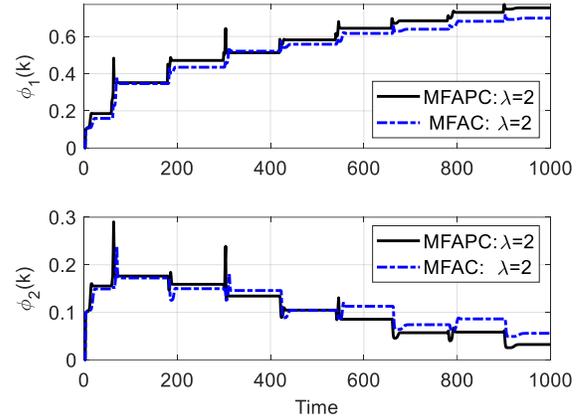

Fig. 3 Estimated value of PG

From Fig. 3, we can see that there is a modest difference in the estimation of PG parameters between MFAPC and MFAC, although all the controller's parameters are set identically. Since the controller design of MFAPC and MFAC merely use the system I/O data, which contains all the information of the system dynamics. The different outputs of both control methods will cause different I/O data, which further leads to a modest difference of the estimated PG results.

## V. CONCLUSION

Based on the combination of the predictive control and MFAC, we propose a novel FFDL-MFAPC method with adjustable parameters for a class of discrete-time SISO nonlinear systems. Then, we show some special cases of MFAPC. The BIBO stability analysis and the monotonic convergence of the tracking error of the MFAPC method are analyzed by the contraction mapping technique with some other lemmas and theorems. The simulations were carried out to verify the effectiveness of the proposed MFAPC. All the results in this paper can be extended to MIMO nonlinear discrete-time systems and we would like to show them in another paper.

## APPENDIX

Proof of Theorem 2. This section shows the proof of convergence of the tracing error and the BIBO stability of the system controlled by proposed MFAPC.

We define $\boldsymbol{P} = \boldsymbol{g}^T[\hat{\boldsymbol{\Psi}}_{Nu}^T(k)\hat{\boldsymbol{\Psi}}_{Nu}(k) + \lambda \boldsymbol{I}]^{-1}\hat{\boldsymbol{\Psi}}_{Nu}^T(k)$, and from Section II, we know that $\hat{\boldsymbol{\Psi}}_U(k)$, $\hat{\boldsymbol{\Psi}}_Y(k)$ can be expressed as

$$\hat{\boldsymbol{\Psi}}_U(k) = \left[\hat{\boldsymbol{\Psi}}_{U1}(k), \hat{\boldsymbol{\Psi}}_{U2}(k), \cdots, \hat{\boldsymbol{\Psi}}_{ULu-1}(k), 0\right]_{N \times Lu},$$

$$\hat{\boldsymbol{\Psi}}_Y(k) = \left[\hat{\boldsymbol{\Psi}}_{Y1}(k), \hat{\boldsymbol{\Psi}}_{Y2}(k), \cdots, \hat{\boldsymbol{\Psi}}_{YLy}(k)\right]_{N \times Ly}$$

We define $\Delta \mathbf{G}(k) = \begin{bmatrix} \Delta u(k) \\ \vdots \\ \Delta u(k-L_u+1) \\ \Delta y(k) \\ \vdots \\ \Delta y(k-L_y+1) \end{bmatrix}$, $\mathbf{F} = \begin{bmatrix} 1 \\ 0 \\ \vdots \\ 0 \end{bmatrix}_{L_y+L_u}$

Then we have

$\Delta \mathbf{G}(k) = [\Delta u(k), \cdots, \Delta u(k-L_u+1), \Delta y(k), \cdots, \Delta y(k-L_y+1)]^T$

$= \begin{bmatrix} \mathbf{g}^T [\tilde{\boldsymbol{\Psi}}_{Nu}^T(k)\tilde{\boldsymbol{\Psi}}_{Nu}(k)+\lambda \mathbf{I}]^{-1}\tilde{\boldsymbol{\Psi}}_3^T(k)[\rho_{L_y+1}(\mathbf{Y}_N^*(k+1) \\ -\mathbf{E}y(k)) - \tilde{\boldsymbol{\Psi}}_Y(k)\Lambda_Y \Delta \mathbf{Y}_{L_y}(k) - \tilde{\boldsymbol{\Psi}}_U(k)\Lambda_U \Delta \mathbf{U}_{L_u}(k-1)] \\ \vdots \\ \Delta u(k-L_u+1) \\ \Delta y(k) \\ \vdots \\ \Delta y(k-L_y+1) \end{bmatrix}$

$= \mathbf{A}(k)\mathbf{K}_1$
$+ \rho_{L_y+1}\mathbf{g}^T[\tilde{\boldsymbol{\Psi}}_{Nu}^T(k)\tilde{\boldsymbol{\Psi}}_{Nu}(k)+\lambda \mathbf{I}]^{-1}\tilde{\boldsymbol{\Psi}}_{Nu}^T(k)\mathbf{E}_N \mathbf{F}e(k)$

(20)

where $\mathbf{K}_1 = [\Delta u(k-1), \cdots, \Delta u(k-L_u+1), \Delta y(k), \cdots,$
$\Delta y(k-L_y+1), \Delta u(k-L_u)]^T$,
$= \mathbf{CD}(k-1)\Delta \mathbf{G}(k-1)$

$\mathbf{A}(k)_{(L_y+L_u)\times(L_y+L_u)} = $
$\begin{bmatrix} -\rho_{L_y+2}\mathbf{P}\hat{\tilde{\boldsymbol{\Psi}}}_{U1} & \cdots & -\rho_{L_y+L_u}\mathbf{P}\hat{\tilde{\boldsymbol{\Psi}}}_{UL_u-1} & -\rho_1 \mathbf{P}\hat{\tilde{\boldsymbol{\Psi}}}_{Y1} & \cdots & -\rho_{L_y}\mathbf{P}\hat{\tilde{\boldsymbol{\Psi}}}_{YL_y} & 0 \\ 1 & & & & & & \\ 0 & \ddots & 0 & & & & \\ & & 1 & \ddots & & & \\ & & & \ddots & 0 & & \\ & & & & \ddots & \ddots & \\ & & & & & 1 & 0 \end{bmatrix}$

$\mathbf{C} = \begin{bmatrix} 1 & & & & & & & \\ & \ddots & & & & & & \\ & & 1 & & & & & \\ & & & 0 & 1 & & & \\ & & & 1 & & & & \\ & & & & & \ddots & 0 & \\ & & & & & 1 & 0 & 0 \end{bmatrix}_{(L_y+L_u)\times(L_y+L_u)}$ $\mathbf{E}_N = \begin{bmatrix} 1 \\ \vdots \\ 1 \end{bmatrix}_{N\times 1}$

$\mathbf{D}(k-1)$
$= \begin{bmatrix} 1 & & & & & & \\ & \ddots & & & & & \\ & & 1 & & & & \\ & & & 1 & & & \\ & & & & \ddots & & \\ & & & & & 1 & \\ \phi_{L_y+1} & \cdots & \phi_{L_y+L_u} & \phi_1 & \cdots & \phi_{L_y-1} & \phi_{L_y} \end{bmatrix}_{(L_y+L_u)\times(L_y+L_u)}$

(20) can be written as
$\Delta \mathbf{G}(k) = \mathbf{A}(k)\mathbf{CD}(k-1)\Delta \mathbf{G}(k-1)$
$\quad + \rho_{L_y+1}\mathbf{g}^T[\hat{\tilde{\boldsymbol{\Psi}}}_{Nu}^T(k)\hat{\tilde{\boldsymbol{\Psi}}}_{Nu}(k)+\lambda \mathbf{I}]^{-1}\hat{\tilde{\boldsymbol{\Psi}}}_{Nu}^T(k)\mathbf{E}_N \mathbf{F}e(k)$
(21)

Considering the sum of the first row of $\mathbf{A}(k)$ and the matrix norm inequalities between $\|\bullet\|_\infty$ and $\|\bullet\|_2$, we have

$\sum_{i=1}^{L_u-1}\left|\rho_{L_y+i+1}\mathbf{P}\hat{\tilde{\boldsymbol{\Psi}}}_{Ui}(k)\right| + \sum_{i=1}^{L_y}\left|\rho_{Li}\mathbf{P}\hat{\tilde{\boldsymbol{\Psi}}}_{Yi}(k)\right|$

$\leq (\max_{i=1,\cdots,L_y+L_u}\rho_i)\left[\sum_{i=1}^{L_u-1}\left|\mathbf{P}\hat{\tilde{\boldsymbol{\Psi}}}_{Ui}(k)\right| + \sum_{i=1}^{L_y}\left|\mathbf{P}\hat{\tilde{\boldsymbol{\Psi}}}_{Yi}(k)\right|\right]$

$\leq (\max_{i=1,\cdots,L_y+L_u}\rho_i)\left\|[\hat{\tilde{\boldsymbol{\Psi}}}_{Nu}^T(k)\hat{\tilde{\boldsymbol{\Psi}}}_{Nu}(k)+\lambda \mathbf{I}]^{-1}\hat{\tilde{\boldsymbol{\Psi}}}_{Nu}^T(k)\right.$

$\quad \bullet [\hat{\tilde{\boldsymbol{\Psi}}}_Y(k), \hat{\tilde{\boldsymbol{\Psi}}}_U(k)]\Big\|_\infty$

$\leq (\max_{i=1,\cdots,L_y+L_u}\rho_i)\sqrt{N_u}\left\|[\hat{\tilde{\boldsymbol{\Psi}}}_{Nu}^T(k)\hat{\tilde{\boldsymbol{\Psi}}}_{Nu}(k)+\lambda \mathbf{I}]^{-1}\right\|_2 \left\|\hat{\tilde{\boldsymbol{\Psi}}}_{Nu}^T(k)\right\|_\infty$

$\quad \bullet \left\|[\hat{\tilde{\boldsymbol{\Psi}}}_Y(k), \hat{\tilde{\boldsymbol{\Psi}}}_U(k)]\right\|_\infty$

(22)

$\hat{\tilde{\boldsymbol{\Psi}}}_{Nu}^T(k)\hat{\tilde{\boldsymbol{\Psi}}}_{Nu}(k)$ is a symmetric semi-positive matrix, which means that $\hat{\tilde{\boldsymbol{\Psi}}}_{Nu}^T(k)\hat{\tilde{\boldsymbol{\Psi}}}_{Nu}(k)+\lambda \mathbf{I}$ will be a symmetric positive matrix, then we have
$\left[\left(\hat{\tilde{\boldsymbol{\Psi}}}_{Nu}^T(k)\hat{\tilde{\boldsymbol{\Psi}}}_{Nu}(k)+\lambda \mathbf{I}\right)^{-1}\right]^T = \left(\hat{\tilde{\boldsymbol{\Psi}}}_{Nu}^T(k)\hat{\tilde{\boldsymbol{\Psi}}}_{Nu}(k)+\lambda \mathbf{I}\right)^{-1}$. $\|\bullet\|_\infty$ is the maximum row sum matrix norm (max norm). $\|\bullet\|_2$ is the spectral norm of matrix. We suppose the eigenvalues of $\hat{\tilde{\boldsymbol{\Psi}}}_{Nu}^T(k)\hat{\tilde{\boldsymbol{\Psi}}}_{Nu}(k)$ are $b_i \geq 0$, $i=1,\cdots,N_u$, so the eigenvalues of $\hat{\tilde{\boldsymbol{\Psi}}}_{Nu}^T(k)\hat{\tilde{\boldsymbol{\Psi}}}_{Nu}(k)+\lambda \mathbf{I}$ are $\lambda+b_i > 0$, $i=1,\cdots,N_u$, which means that the eigenvalues of $[\hat{\tilde{\boldsymbol{\Psi}}}_{Nu}^T(k)\hat{\tilde{\boldsymbol{\Psi}}}_{Nu}(k)+\lambda \mathbf{I}]^{-1}$ are $\frac{1}{\lambda+b_i} > 0$, $i=1,\cdots,N_u$. Therefore, we get

$\left\|[\hat{\tilde{\boldsymbol{\Psi}}}_{Nu}^T(k)\hat{\tilde{\boldsymbol{\Psi}}}_{Nu}(k)+\lambda \mathbf{I}]^{-1}\right\|_2$

$= \sqrt{\sigma\left(\left([\hat{\tilde{\boldsymbol{\Psi}}}_{Nu}^T(k)\hat{\tilde{\boldsymbol{\Psi}}}_{Nu}(k)+\lambda \mathbf{I}]^{-1}\right)^T [\hat{\tilde{\boldsymbol{\Psi}}}_{Nu}^T(k)\hat{\tilde{\boldsymbol{\Psi}}}_{Nu}(k)+\lambda \mathbf{I}]^{-1}\right)}$

$= \sqrt{\sigma\left(\left([\hat{\tilde{\boldsymbol{\Psi}}}_{Nu}^T(k)\hat{\tilde{\boldsymbol{\Psi}}}_{Nu}(k)+\lambda \mathbf{I}]^{-1}\right)^2\right)} = \frac{1}{\min_{i=1,\cdots Nu}\{\lambda+b_i\}}$

(23)

Combining (22) and (23), we have

$\sum_{i=1}^{L_u-1}\left|\mathbf{P}\hat{\tilde{\boldsymbol{\Psi}}}_{Ui}(k)\right| + \sum_{i=1}^{L_y}\left|\mathbf{P}\hat{\tilde{\boldsymbol{\Psi}}}_{Yi}(k)\right|$

$\leq \sqrt{N_u}\left\|[\hat{\tilde{\boldsymbol{\Psi}}}_{Nu}^T(k)\hat{\tilde{\boldsymbol{\Psi}}}_{Nu}(k)+\lambda \mathbf{I}]^{-1}\right\|_2 \left\|\hat{\tilde{\boldsymbol{\Psi}}}_{Nu}^T(k)\right\|_\infty \left\|[\hat{\tilde{\boldsymbol{\Psi}}}_Y(k), \hat{\tilde{\boldsymbol{\Psi}}}_U(k)]\right\|_\infty$

$\leq \sqrt{N_u}\frac{1}{\min_{i=1,\cdots Nu}\{\lambda+b_i\}}\left\|\hat{\tilde{\boldsymbol{\Psi}}}_{Nu}^T(k)\right\|_\infty \left\|[\hat{\tilde{\boldsymbol{\Psi}}}_Y(k), \hat{\tilde{\boldsymbol{\Psi}}}_U(k)]\right\|_\infty$

(24)

Assume $s$ is the number of the maximum sum of the row of $\hat{\tilde{\boldsymbol{\Psi}}}_{Nu}^T(k)$, then we can see that

$$\left\|\hat{\tilde{\boldsymbol{\Psi}}}_{Nu}^T(k)\right\|_\infty = \left|\sum_{m=s}^{N}\sum_{j=s}^{m}[\hat{\boldsymbol{\phi}}_{Lu}^T(k+j-1)\boldsymbol{A}^{j-s}\boldsymbol{B} + \hat{\boldsymbol{\phi}}_{Ly}^T(k+j-1)\sum_{i=0}^{j-s-1}\boldsymbol{C}^i\boldsymbol{D}\psi_{j-i-1,s}]\right|$$

is bounded. We suppose that $s_1$ is the number of the maximum sum of the row of the matrix $[\hat{\tilde{\boldsymbol{\Psi}}}_Y(k),\hat{\tilde{\boldsymbol{\Psi}}}_U(k)]$, then we can see that

$$\left\|[\hat{\tilde{\boldsymbol{\Psi}}}_Y(k),\hat{\tilde{\boldsymbol{\Psi}}}_U(k)]\right\|_\infty = \sum_{j=1}^{s_1}[\hat{\boldsymbol{\phi}}_{Ly}^T(k+j-1)\boldsymbol{C}^{j-1} + \hat{\boldsymbol{\phi}}_{Ly}^T(k+j-1)\sum_{i=0}^{j-2}\boldsymbol{C}^i\boldsymbol{D}\hat{\varphi}_{1,j-i-1}]$$
$$+\sum_{j=1}^{s_1}[\hat{\boldsymbol{\phi}}_{Lu}^T(k+j-1)\boldsymbol{A}^j + \hat{\boldsymbol{\phi}}_{Lu}^T(k+j-1)\sum_{i=0}^{j-2}\boldsymbol{C}^i\boldsymbol{D}\hat{\varphi}_{2,j-i-1}]$$

is bounded. Therefore, there exists a positive $\lambda_{\min 1}$, such that $\lambda > \lambda_{\min 1}$, we can obtain the following inequation:

$$\left[\sum_{i=1}^{Ly}\left|\boldsymbol{P}\hat{\tilde{\boldsymbol{\Psi}}}_{Yi}(k)\right| + \sum_{i=1}^{Lu-1}\left|\boldsymbol{P}\hat{\tilde{\boldsymbol{\Psi}}}_{Ui}(k)\right|\right]^{\frac{1}{Ly+Lu-1}}$$
$$\leq \left[\sqrt{N_u}\frac{1}{\min_{i=1,\cdots Nu}\{\lambda+b_i\}}\left\|\hat{\tilde{\boldsymbol{\Psi}}}_{Nu}^T(k)\right\|_\infty \left\|[\hat{\tilde{\boldsymbol{\Psi}}}_Y(k),\hat{\tilde{\boldsymbol{\Psi}}}_U(k)]\right\|_\infty\right]^{\frac{1}{Ly+Lu-1}}$$
$$\leq M_1 < 1$$
(25)

Given $0 < \rho_1 < 1$, $\cdots$, $0 < \rho_{Ly+Lu} < 1$, we have $(\max_{i=1,\cdots,Ly+Lu}\rho_i) < 1$. Hence, we have

$$\sum_{i=1}^{Lu-1}\left|\rho_{Ly+1+i}\boldsymbol{P}\hat{\tilde{\boldsymbol{\Psi}}}_{Ui}(k)\right| + \sum_{i=1}^{Ly}\left|\rho_i\boldsymbol{P}\hat{\tilde{\boldsymbol{\Psi}}}_{Yi}(k)\right|$$
$$\leq (\max_{i=1,\cdots,Ly+Lu}\rho_i)\left[\sum_{i=1}^{Lu-1}\left|\boldsymbol{P}\hat{\tilde{\boldsymbol{\Psi}}}_{Ui}(k)\right| + \sum_{i=1}^{Ly}\left|\boldsymbol{P}\hat{\tilde{\boldsymbol{\Psi}}}_{Yi}(k)\right|\right] \quad (26)$$
$$\leq (\max_{i=1,\cdots,Ly+Lu}\rho_i)M_1^{Ly+Lu-1} < 1$$

According to Lemma 1 and (26), the sum of the absolute values of each element in the first raw of matrix $\boldsymbol{A}(k)$ is less than 1. We can see that all the eigenvalues of $\boldsymbol{A}(k)$ satisfy $|z| < 1$. The characteristic equation of $\boldsymbol{A}(k)$ is

$$z^{Ly+Lu} + \rho_{Ly+2}\boldsymbol{P}\tilde{\boldsymbol{\Psi}}_{U1}z^{Ly+Lu-1} + \cdots + \rho_{Ly+Lu}\boldsymbol{P}\tilde{\boldsymbol{\Psi}}_{ULu-1}z^{Ly+1}$$
$$+ \rho_1\boldsymbol{P}\tilde{\boldsymbol{\Psi}}_{Y1}z^{Ly} + \cdots + \rho_{Ly}\boldsymbol{P}\tilde{\boldsymbol{\Psi}}_{YLy}z = 0 \quad (27)$$

Given $|z| < 1$ and (27), we have the following inequation:

$$|z|^{Ly+Lu-1} \leq (\max_{i=1,\cdots,Ly+Lu}\rho_i)\left[\sum_{i=1}^{Lu-1}\left|\boldsymbol{P}\hat{\tilde{\boldsymbol{\Psi}}}_{Ui}(k)\right| + \sum_{i=1}^{Ly}\left|\boldsymbol{P}\hat{\tilde{\boldsymbol{\Psi}}}_{Yi}(k)\right|\right]$$
$$\leq (\max_{i=1,\cdots,Ly+Lu}\rho_i)M_1^{Ly+Lu-1} < 1$$
(28)

which means $|z| \leq (\max_{i=1,\cdots,Ly+Lu}\rho_i)^{1/Ly+Lu-1}M_1 < 1$. Therefore, according to Lemma 2 and (28), there exists an arbitrarily small positive $\varepsilon$ that makes the following inequation hold.

$$|\boldsymbol{A}(k)|_v \leq s(\boldsymbol{A}(k)) + \varepsilon \leq (\max_{i=1,\cdots,Ly+Lu}\rho_i)^{1/Ly+Lu-1}M_1 + \varepsilon < 1 \quad (29)$$

where $|\boldsymbol{A}(k)|_v$ is the compatible norm of $\boldsymbol{A}(k)$. Let $d_1 = (\max_{i=1,\cdots,Ly+Lu}\rho_i)^{1/Ly+Lu-1}M_1$.

Then, according to the definition of spectral radius and Lemma 2, [14] has deduced the following inequation

$$\|\boldsymbol{A}(k)\|_v\|\boldsymbol{C}\|_v\|\boldsymbol{D}(k-1)\|_v$$
$$\leq (d_1+\varepsilon)(1+\varepsilon)(\max(1,b)+\varepsilon) \triangleq d_2 < 1 \quad (30)$$

Here, $\boldsymbol{PE}_N = \boldsymbol{g}^T[\hat{\tilde{\boldsymbol{\Psi}}}_{Nu}^T(k)\hat{\tilde{\boldsymbol{\Psi}}}_{Nu}(k) + \lambda\boldsymbol{I}]^{-1}\hat{\tilde{\boldsymbol{\Psi}}}_{Nu}^T(k)\boldsymbol{E}_N$ is obviously a number which equals to the sum of the first row of $[\hat{\tilde{\boldsymbol{\Psi}}}_{Nu}^T(k)\hat{\tilde{\boldsymbol{\Psi}}}_{Nu}(k) + \lambda\boldsymbol{I}]^{-1}\hat{\tilde{\boldsymbol{\Psi}}}_{Nu}^T(k)$. Then we have

$$\boldsymbol{g}^T[\hat{\tilde{\boldsymbol{\Psi}}}_{Nu}^T(k)\hat{\tilde{\boldsymbol{\Psi}}}_{Nu}(k) + \lambda\boldsymbol{I}]^{-1}\hat{\tilde{\boldsymbol{\Psi}}}_{Nu}^T(k)\boldsymbol{E}_N$$
$$\leq \left\|[\hat{\tilde{\boldsymbol{\Psi}}}_{Nu}^T(k)\hat{\tilde{\boldsymbol{\Psi}}}_{Nu}(k) + \lambda\boldsymbol{I}]^{-1}\hat{\tilde{\boldsymbol{\Psi}}}_{Nu}^T(k)\right\|_\infty$$
$$\leq \sqrt{N_u}\left\|[\hat{\tilde{\boldsymbol{\Psi}}}_{Nu}^T(k)\hat{\tilde{\boldsymbol{\Psi}}}_{Nu}(k) + \lambda\boldsymbol{I}]^{-1}\right\|_2 \left\|\hat{\tilde{\boldsymbol{\Psi}}}_{Nu}^T(k)\right\|_\infty \quad (31)$$
$$\leq \sqrt{N_u}\frac{1}{\min_{i=1,\cdots Nu}\{\lambda+b_i\}}\left\|\hat{\tilde{\boldsymbol{\Psi}}}_{Nu}^T(k)\right\|_\infty$$

Similar to the proof process of (25), there exists positive $\lambda_{\min 2}$ and $M_2$, such that $\lambda > \lambda_{\min 2}$, then we have the following two inequations

$$0 < \boldsymbol{g}^T\left[\hat{\tilde{\boldsymbol{\Psi}}}_{Nu}^T(k)\hat{\tilde{\boldsymbol{\Psi}}}_{Nu}(k) + \lambda\boldsymbol{I}\right]^{-1}\hat{\tilde{\boldsymbol{\Psi}}}_{Nu}^T(k)\boldsymbol{E}_N \leq M_2 < 1 \quad (32)$$
$$d_3 \triangleq \rho_{Ly+1}M_2\left\|\boldsymbol{X}\boldsymbol{\phi}_L^T(i+1)\right\|_v < 0.5 \quad (33)$$

where, $\boldsymbol{X} = \begin{bmatrix} \boldsymbol{I}_{Lu} \\ \boldsymbol{I}_{Ly} \end{bmatrix}$

Taking the norm of (20) and combining (30) and (32), we have

$$\|\Delta\boldsymbol{G}(k)\|_v = \|\boldsymbol{A}(k)\|_v\|\boldsymbol{C}\|_v\|\boldsymbol{D}(k-1)\|_v\|\Delta\boldsymbol{G}(k-1)\|_v$$
$$+ \left|\rho_{Ly+1}\boldsymbol{g}^T[\tilde{\boldsymbol{\Psi}}_{Nu}^T(k)\tilde{\boldsymbol{\Psi}}_{Nu}(k) + \lambda\boldsymbol{I}]^{-1}\tilde{\boldsymbol{\Psi}}_{Nu}^T(k)\boldsymbol{E}_N\right||e(k)|$$
$$= d_2\|\Delta\boldsymbol{G}(k-1)\|_v + M_2|e(k)|$$
$$\vdots$$
$$= d_2^k\|\Delta\boldsymbol{G}(0)\|_v + \rho_{Ly+1}M_2\sum_{i=1}^{k}d_2^{k-i}|e(i)|$$
(34)

Combining (3), (12), (13) and (20) together, we have

$$e(k+1) = y^* - y(k+1) = y^* - y(k) - \boldsymbol{\phi}_L^T(k)\Delta\boldsymbol{H}(k)$$
$$= e(k) - \boldsymbol{X}\boldsymbol{\phi}_L^T(k)[\boldsymbol{A}(k)\boldsymbol{C}\boldsymbol{D}(k-1)\Delta\boldsymbol{G}(k-1)$$
$$+ \rho_{Ly+1}\boldsymbol{g}^T[\hat{\tilde{\boldsymbol{\Psi}}}_{Nu}^T(k)\hat{\tilde{\boldsymbol{\Psi}}}_{Nu}(k) + \lambda\boldsymbol{I}]^{-1}\hat{\tilde{\boldsymbol{\Psi}}}_{Nu}^T(k)\boldsymbol{E}_N\boldsymbol{F}e(k)]$$
$$= (1 - \rho_{Ly+1}\boldsymbol{\phi}_{Ly+1}(k)\boldsymbol{g}^T[\hat{\tilde{\boldsymbol{\Psi}}}_{Nu}^T(k)\hat{\tilde{\boldsymbol{\Psi}}}_{Nu}(k) + \lambda\boldsymbol{I}]^{-1}$$
$$\bullet \hat{\tilde{\boldsymbol{\Psi}}}_{Nu}^T(k)\boldsymbol{E}_N)e(k) - \boldsymbol{X}\boldsymbol{\phi}_L^T(k)\boldsymbol{A}(k)\boldsymbol{C}\boldsymbol{D}(k-1)\Delta\boldsymbol{G}(k-1)$$
$$= (1 - \rho_{Ly+1}\boldsymbol{\phi}_{Ly+1}(k)\boldsymbol{P})e(k)$$
$$- \boldsymbol{X}\boldsymbol{\phi}_L^T(k)\boldsymbol{A}(k)\boldsymbol{C}\boldsymbol{D}(k-1)\Delta\boldsymbol{G}(k-1)$$
(35)

Similarly, there exists a positive $\lambda_{\min 3}$ and a positive $M_3$, such that $\lambda > \lambda_{\min 3}$, then we have the below inequation

$$0 < M_3 \leq |\rho_{Ly+1}\phi_{Ly+1}(k)\boldsymbol{P}|$$
$$= |\rho_{Ly+1}\phi_{Ly+1}(k)\boldsymbol{g}^T[\hat{\boldsymbol{\Psi}}_{Nu}^T(k)\hat{\boldsymbol{\Psi}}_{Nu}(k) + \lambda \boldsymbol{I}]^{-1}\hat{\boldsymbol{\Psi}}_{Nu}^T(k)\boldsymbol{E}_N|  \quad (36)$$
$$\leq \rho_{Ly+1}|\phi_{Ly+1}(k)|\frac{1}{\min_{i=1,\cdots Nu}\{\lambda + b_i\}}\|\hat{\boldsymbol{\Psi}}_{Nu}(k)\|_\infty < 0.5$$

According to (36), we have
$$0.5 < |1 - \rho_{Ly+1}\phi_{Ly+1}(k)\boldsymbol{P}| \leq |1 - \rho_{Ly+1}|\phi_{Ly+1}(k)\boldsymbol{P}| \leq 1 - M_3 < 1 \quad (37)$$

Let $d_4 = 1 - M_3$ and take the norm of (34), then we yield
$$|e(k+1)| = |1 - \rho_{Ly+1}\phi_{Ly+1}(k)\boldsymbol{P}||e(k)|$$
$$+ \|\boldsymbol{X}\boldsymbol{\phi}_L^T(k)\|_v \|\boldsymbol{A}(k)\boldsymbol{C}\boldsymbol{D}(k-1)\|_v \|\Delta \boldsymbol{G}(k-1)\|_v$$
$$< d_4|e(k)| + d_2\|\boldsymbol{X}\boldsymbol{\phi}_L^T(k)\|_v \|\Delta \boldsymbol{G}(k-1)\|_v < \cdots$$
$$< d_4^{k-1}|e(2)| + d_2\sum_{i=1}^{k-1}d_4^{k-1-i}\|\boldsymbol{X}\boldsymbol{\phi}_L^T(i+1)\|_v \|\Delta \boldsymbol{G}(k-1)\|_v$$
$$< d_4^{k-1}|e(2)| + d_2^i\|\Delta \boldsymbol{G}(0)\|_v$$
$$+ d_2\sum_{i=1}^{k-1}d_4^{k-1-i}\|\boldsymbol{X}\boldsymbol{\phi}_L^T(i+1)\|_v [\rho_{Ly+1}M_2\sum_{j=1}^{i}d_2^{i-j}|e(j)|] \quad (38)$$

Then (38) becomes
$$|e(k+1)| < d_4^{k-1}|e(2)| + d_2 d_3\sum_{i=1}^{k-1}d_4^{k-1-i}\sum_{j=1}^{i}d_2^{i-j}|e(j)|$$
$$+ (d_4^{k-2} + d_4^{k-3}d_2 + \cdots + d_2^{k-2})d_2^2 b \|\Delta \boldsymbol{G}(0)\|_v \quad (39)$$

Let
$$g(k+1) = d_4^{k-1}|e(2)| + d_2 d_3\sum_{i=1}^{k-1}d_4^{k-1-i}\sum_{j=1}^{i}d_2^{i-j}|e(j)|$$
$$+ (d_4^{k-2} + d_4^{k-3}d_2 + \cdots + d_2^{k-2})d_2^2 b \|\Delta \boldsymbol{G}(0)\|_v \quad (40)$$

Inequation (39) can be rewritten as follow
$$|e(k+1)| < g(k+1), \quad k=1, 2, \ldots  \quad (41)$$

Where, $g(2) = d_3|e(1)|$, $d_4 = 1 - M_3 > 0.5 > d_3$

According to (40) and (41), we have
$$g(k+2) = d_4^k|e(2)| + d_2 d_3\sum_{i=1}^{k}d_4^{k-i}\sum_{j=1}^{i}d_2^{i-j}|e(j)|$$
$$+ (d_4^{k-1} + d_4^{k-3}d_2 + \cdots + d_2^{k-1})d_2^2 b \|\Delta \boldsymbol{G}(0)\|_v$$
$$= d_4 g(k+1) + d_2 d_3\sum_{j=1}^{k-1}d_2^{k-j}|e(j)| + d_2 d_3|e(k)| + d_2^{k+1}b\|\Delta \boldsymbol{G}(0)\|_v$$
$$\leq d_4 g(k+1) + d_2 d_3\sum_{j=1}^{k-1}d_2^{k-j}|e(j)| + d_2 d_3 g(k) + d_2^{k+1}b\|\Delta \boldsymbol{G}(0)\|_v \quad (42)$$

Let
$$h(k) = d_2 d_3\sum_{j=1}^{k-1}d_2^{k-j}|e(j)| + d_2 d_3 g(k) + d_2^{k+1}b\|\Delta \boldsymbol{G}(0)\|_v \quad (43)$$

From above analysis, we have
$$d_3 = \rho_{Ly+1}M_2\|\boldsymbol{X}\boldsymbol{\phi}_L^T(i+1)\|_v < 0.5 < d_4 < 1 \quad (44)$$

From (43) combined with (44), we have

$$h(k) < d_2 d_3\sum_{j=1}^{k-1}d_2^{k-j}|e(j)| + d_2 d_4 g(k) + d_2^{k+1}b\|\Delta \boldsymbol{G}(0)\|_v$$
$$= d_2 d_3\sum_{j=1}^{k-1}d_2^{k-j}|e(j)| + d_2^{k+1}b\|\Delta \boldsymbol{G}(0)\|_v + d_2 d_4[d_4^{k-2}|e(2)|$$
$$+ d_2 d_3[\sum_{i=1}^{k-2}d_4^{k-2-i}\sum_{j=1}^{i}d_2^{i-j}|e(j)|$$
$$+ (d_4^{k-3} + d_4^{k-4}d_2 + \cdots + d_2^{k-3})d_2^2 b \|\Delta \boldsymbol{G}(0)\|_v]$$
$$= d_2 g(k+1) \quad (45)$$

Substituting (45) into (41), we get
$$g(k+2) < (d_4 + d_2)g(k+1) \quad (46)$$

When $0 < \rho_i < 1$, $(i=1,2,\cdots,L_y+L_u)$, we can get $0 < (\max_{i=1,\cdots,Ly+Lu}\rho_i)^{1/Ly+Lu-1}M_1 < M_3 < 1$, we further yield
$$d_4 + d_2 = 1 - M_3 + (\max_{i=1,\cdots,Ly+Lu}\rho_i)^{1/Ly+Lu-1}M_1 < 1 \quad (47)$$

Substituting (47) into (46), we have
$$\lim_{k\to\infty}g(k+2) < \lim_{k\to\infty}(d_4+d_2)g(k+1) < \cdots < \lim_{k\to\infty}(d_4+d_2)^k g(2) = 0 \quad (48)$$

Thus, *Theorem 2* is the direct result of (48) and (41) if $\lambda > \lambda_{\min} = \max\{\lambda_{\min 1}, \lambda_{\min 2}, \lambda_{\min 3}\}$.

Since $\boldsymbol{G}(k)$ is the information vector that consists of the inputs and outputs, we can prove the BIBO stability of the closed loop system by proving the boundedness of $\boldsymbol{G}(k)$.

From (34), (40), (41) and (46), we have
$$\|\boldsymbol{G}(k)\|_v \leq \sum_{i=0}^{k}\|\Delta \boldsymbol{G}(i)\|_v$$
$$\leq \sum_{j=1}^{k}\left[d_2^j\|\Delta \boldsymbol{G}(0)\|_v + \rho_{Ly+1}M_2\sum_{i=1}^{j}d_2^{j-i}|e(i)|\right]$$
$$< \frac{\|\Delta \boldsymbol{G}(0)\|_v}{1-d_2} + \rho_{Ly+1}M_2\sum_{j=1}^{k}\sum_{i=1}^{j}d_2^{j-i}|e(i)|$$
$$= \frac{\|\Delta \boldsymbol{G}(0)\|_v}{1-d_2} + \rho_{Ly+1}M_2[(|e(1)| + (d_2^1|e(1)| + |e(2)|)$$
$$+ (d_2^2|e(1)| + d_2^1|e(2)| + |e(3)|) + \cdots$$
$$+ (d_2^{k-1}|e(1)| + \cdots + |e(k)|)]$$
$$< \frac{\|\Delta \boldsymbol{G}(0)\|_v}{1-d_2} + \frac{\rho_{Ly+1}M_2}{1-d_2}(|e(1)| + |e(2)| + \cdots + |e(k)|)$$
$$< \frac{\|\Delta \boldsymbol{G}(0)\|_v}{1-d_2} + \frac{\rho_{Ly+1}M_2}{1-d_2}(|g(1)| + |g(2)| + \cdots + |g(k)|)$$
$$< \frac{\|\Delta \boldsymbol{G}(0)\|_v}{1-d_2} + \frac{\rho_{Ly+1}M_2}{1-d_2}\frac{|g(2)|}{1-d_2-d_4} \quad (49)$$

Therefore, the boundedness of $\|\boldsymbol{G}(k)\|_v$ is proved by (49). In other words, the closed-loop system is BIBO stable.
We finished the proof of *Theorem* 2.

## References

[1] Y. J. Liu, S. Tong, C. L. P Chen, D. J. Li. Neural controller design-based adaptive control for nonlinear MIMO systems with unknown hysteresis inputs[J]. IEEE transactions on cybernetics, 2015, 46(1): 9-19.

[2] Liu Y J, Tong S. Barrier Lyapunov functions for Nussbaum gain adaptive control of full state constrained nonlinear systems[J]. Automatica, 2017, 76: 143-152.

[3] Pekař L, Matušů R. A suboptimal shifting based zero-pole placement method for systems with delays[J]. International Journal of Control, Automation and Systems, 2018, 16(2): 594-608.

[4] Kugelmann B, Pulch R. Robust Optimal Control of Fishing in a Three Competing Species Model[J]. IFAC-PapersOnLine, 2018, 51(2): 7-12.

[5] Li Y X, Yang G H. Model-based adaptive event-triggered control of strict-feedback nonlinear systems[J]. IEEE transactions on neural networks and learning systems, 2018, 29(4): 1033-1045.

[6] Silva L R, Flesch R C C, Normey-Rico J E. Controlling industrial dead-time systems: When to use a PID or an advanced controller[J]. ISA transactions, 2019.

[7] Kalman R E. Design of self-optimizing control system[J]. Trans. ASME, 1958, 80: 468-478.

[8] Lei W, Li C, Chen M Z Q. Robust adaptive tracking control for quadrotors by combining PI and self-tuning regulator[J]. IEEE Transactions on Control Systems Technology, 2018.

[9] Åström K J, Wittenmark B. Adaptive control[M]. Courier Corporation, 2013.

[10] Yang C, Jiang Y, He W, et al. Adaptive parameter estimation and control design for robot manipulators with finite-time convergence[J]. IEEE Transactions on Industrial Electronics, 2018, 65(10): 8112-8123.

[11] Grema A S, Cao Y. Dynamic Self-Optimizing Control for Uncertain Oil Reservoir Waterflooding Processes[J]. IEEE Transactions on Control Systems Technology, 2019.

[12] Gao W, Jiang Z P, Lewis F L, et al. Leader-to-formation stability of multiagent systems: An adaptive optimal control approach[J]. IEEE Transactions on Automatic Control, 2018, 63(10): 3581-3587.

[13] B. D. O. Anderson and A. Dehghani, "Challenges of adaptive control: Past, permanent and future," Annual reviews in control, vol. 32, no. 2, pp. 123–135, Dec. 2008.

[14] Hou Z, Xiong S. On Model Free Adaptive Control and its Stability Analysis[J]. IEEE Transactions on Automatic Control, 2019.

[15] Hou Z, Jin S. A novel data-driven control approach for a class of discrete-time nonlinear systems[J]. IEEE Transactions on Control Systems Technology, 2011, 19(6): 1549-1558.

[16] Z. S. Wang, L. Liu, and H. G. Zhang, "Neural network-based model free adaptive fault-tolerant control for discrete-time nonlinear systems with sensor fault," IEEE Transaction on Systems Man and Cybernetics: Systems, vol. 47, no. 8,pp. 2351-2362, 2017.

[17] Z. H. Pang, G. P. Liu, D. H. Zhou, and D. H. Sun, "Data-driven control with input design-based data dropout compensation for networked nonlinear systems," IEEE Transaction on Control Systems Technology, vol. 25, no. 2, pp. 628-636, 2017

[18] Y. Zhao, Z. C. Yuan, C. Lu, G. L. Zhang, X. Li, and Y. Chen, "Improved model-free adaptive wide-area coordination damping controller for multiple-input multiple-output power systems," IET Generation, Transmission and Distribution, vol. 10, no. 3, pp. 3264-3275, 2016.

[19] H. G. Zhang, J. G. Zhou, Q. Y. Sun, J. M. Guerrero, and D. Z. Ma, "Data-driven control for interlinked AC/DC microgrids via model-free adaptive control and dual-droop control," IEEE Transactions on Smart Grid, vol. 8, no. 2, pp. 557-571, 2017.

[20] R. C. Roman, M. B. Radac, and R. E. Precup, "Multi-input-multi-output system experimental validation of model-free control and virtual reference feedback tuning techniques," IET Control Theory and Applications, vol. 10, no. 12, pp. 1395-1403, 2016.

[21] Z. S. Hou and S. T. Jin, Model Free Adaptive Control: Theory and Applications, CRC Press, Taylor and Francis Group, 2013

[22] Z. S. Hou and S. T. Jin, "A novel data-driven control approach for a class of discrete-time nonlinear systems," IEEE Transaction on Control Systems Technology, vol. 19, no. 6, pp. 1549-1558, 2011.

[23] Z. S. Hou. On model-free adaptive control: the state of the art and perspective[J]. Control Theory & Applications, 2006, 23(4): 586-592.

[24] Yu Q, Hou Z, Bu X, et al. RBFNN-Based Data-Driven Predictive Iterative Learning Control for Nonaffine Nonlinear Systems[J]. IEEE transactions on neural networks and learning systems, 2019.

[25] Hou Z, Liu S, Tian T. Lazy-learning-based data-driven model-free adaptive predictive control for a class of discrete-time nonlinear systems[J]. IEEE transactions on neural networks and learning systems, 2016, 28(8): 1914-1928.

[26] Hou Z, Liu S, Yin C. Local learning-based model-free adaptive predictive control for adjustment of oxygen concentration in syngas manufacturing industry[J]. IET Control Theory & Applications, 2016, 10(12): 1384-1394.

[27] Dong N, Feng Y, Han X, et al. An Improved Model-free Adaptive Predictive Control Algorithm For Nonlinear Systems With Large Time Delay[C]//2018 IEEE 7th Data Driven Control and Learning Systems Conference (DDCLS). IEEE, 2018: 60-64.

[28] E. I. Jury, Theory and Application of the z-Transform Method. New York: Wiley, 1964.

[29] Abadir K M, Magnus J R. Matrix algebra[M]. Cambridge University

[30] Hou Z, Jin S. Data-driven model-free adaptive control for a class of MIMO nonlinear discrete-time systems[J]. IEEE Transactions on Neural Networks, 2011, 22(12): 2173-2188.

[31] Chongzhi F, Deyun X. Process identification[J]. Beijing: Tsing, 1988.

[32] Hou Z, Zhu Y. Controller-dynamic-linearization-based model free adaptive control for discrete-time nonlinear systems[J]. IEEE Transactions on Industrial Informatics, 2013, 9(4): 2301-2309.